\documentclass{emulateapj}

\usepackage{amsmath}
\usepackage{amssymb}
\usepackage{graphicx}
\usepackage{color}
\usepackage{natbib}
\usepackage{epsfig}

\newcommand{\lFig}[1]{{\label{fig:#1}}}

\newcommand{\djw}[1]{#1}
\newcommand{\ken}[1]{#1}

\def\gtaprx {\lower .1ex\hbox{\rlap{\raise .6ex\hbox{\hskip .3ex
	{\ifmmode{\scriptscriptstyle >}\else
		{$\scriptscriptstyle >$}\fi}}}
	\kern -.4ex{\ifmmode{\scriptscriptstyle \sim}\else
		{$\scriptscriptstyle\sim$}\fi}}}
\def\ltaprx {\lower .1ex\hbox{\rlap{\raise .6ex\hbox{\hskip .3ex
	{\ifmmode{\scriptscriptstyle <}\else
		{$\scriptscriptstyle <$}\fi}}}
	\kern -.4ex{\ifmmode{\scriptscriptstyle \sim}\else
		{$\scriptscriptstyle\sim$}\fi}}}

\newcommand{\cutt}[1]{\textcolor{blue}{}}

\newcommand{\Ms}{{\ensuremath{{M}_{\odot} }}}
\newcommand{\Zs}{\ensuremath{Z_\odot}}

\newcommand{\CASTRO}{\texttt{CASTRO}}
\newcommand{\ZEUS}{\texttt{ZEUS-MP}}

\begin{document}

\title{How the First Stars Regulated Star Formation. II. Enrichment by Nearby Supernovae} 

\author{Ke-Jung Chen\altaffilmark{1,2,3*}, Daniel J. Whalen\altaffilmark{4}, Katharina M. J. 
Wollenberg\altaffilmark{5}, Simon C. O. Glover\altaffilmark{5}, and Ralf S. Klessen\altaffilmark{
5,6}}

\altaffiltext{1}{ Division of Theoretical Astronomy, National Astronomical Observatory of Japan, 
Tokyo 181-8588, Japan} 
\altaffiltext{2}{Institute of Astronomy and Astrophysics, Academia Sinica,  Taipei 10617, Taiwan}
\altaffiltext{3}{Department of Astronomy \& Astrophysics, University of California, Santa Cruz, CA 
95064, USA}
\altaffiltext{4}{Institute of Cosmology and Gravitation, Portsmouth University, Portsmouth, UK}
\altaffiltext{5}{Zentrum f\"{u}r Astronomie, Institut f\"{u}r Theoretische Astrophysik, Universit\"{a}t 
Heidelberg, Germany}
\altaffiltext{6}{Interdisziplin\"{a}res Zentrum f\"{u}r Wissenschaftliches Rechnen, Universit\"{a}t 
Heidelberg, Germany}

\altaffiltext{*}{EACOA Fellow, email: {\tt ken.chen@nao.ac.jp}} 

\begin{abstract}

Metals from Population III (Pop III) supernovae led to the formation of less massive Pop II stars 
in the early universe, altering the course of evolution of primeval galaxies and cosmological 
reionization.  There are a variety of scenarios in which heavy elements from the first supernovae 
were taken up into second-generation stars, but cosmological simulations only model them on the 
largest scales.  We present small-scale, high-resolution simulations of the chemical enrichment of 
a primordial halo by a nearby supernova after partial evaporation by the progenitor star.  We find 
that ejecta from the explosion crash into and mix violently with ablative flows driven off the halo by 
the star, creating dense, enriched clumps capable of collapsing into Pop II stars.  Metals may mix 
less efficiently with the partially exposed core of the halo, so it might form either Pop III or Pop II 
stars. Both Pop II and III stars may thus form after the collision if the ejecta do not strip all the gas 
from the halo.  The partial evaporation of the halo prior to the explosion is crucial to its later 
enrichment by the supernova.

\end{abstract}

  
\section{Introduction}

How metals from the first supernovae (SNe) are taken up into new stars governs the transition from
Pop III to Pop II star formation in the primordial universe \citep{bfcl01,sfno02,mbh03,ss07,clark08,
bsmith09,japp09,dop11,chiaki12,ritt12,ss13,chen15,rit16}.  This transition in turn regulates the luminosities,
spectra and star formation rates of primeval galaxies \citep{get08,jlj09,get10,jeon11,pmb11,wise12,
pmb12,ren15}, the demographics and rates of supermassive black hole formation \citep{th09,awa09,
pm11,wf12,pm12,vol12,agarw12,pm13,choi13,latif13c,latif13a,schl13,jet14,smidt17,tyr17,hle17}, and 
the onset of cosmological reionization \citep{wan04,ket04,abs06,awb07}. This transition also 
influences the rate at which primordial SNe may be detected in future surveys \citep{wet08c,kasen11,met12a,wet12d,wet12a,wet12c,ds13,wet13b,wet13a,wet13c,mes13a,wet13d,wet13e,smidt13a,ds14,chen14c,chen14b,chen14a,smidt14a,magg16}  and the number of low-mass Pop III stars and second-generation stars expected in Galactic 
archaeological searches \citep{bc05,fet05,jw11,til15a}.

Cosmological simulations usually take this transition to occur when metals from Pop III SNe touch a 
halo because they cannot resolve the actual interaction of the metals with the gas in the halo. They 
may therefore overestimate the true mixing efficiency and predict a premature transition from Pop III 
to Pop II star formation. Past, higher resolution simulations seem to indicate that metals from SNe in
reality do not mix with nearby halos very efficiently.  \citet{cr08} modeled SN shocks colliding with 
10$^6$ - 10$^7$ \Ms\ halos at 10 - 100 km s$^{-1}$ at $z \sim$ 10.  They found that mild shear 
flows driven by Kelvin-Helmholtz (KH) instabilities limited mixing to the outer regions of the halo, and 
that at most a third of the baryons at high densities in the halo were enriched above 3\% of the 
metallicity of the ambient intergalactic medium (IGM) by $z \sim$ 6. They concluded that the Pop III / 
Pop II transition may therefore have been protracted, with Pop III stars forming at lower redshifts than
previously thought. More recent studies indicate that metals can reach greater depths in the gas if it is 
turbulent, although some rely on {\it ad hoc} subgrid models for turbulent mixing \citep{gray11,ret13}.

But these models do not include the photoevaporation of the halo by the star prior to its explosion.  
Studies have shown that ionizing UV radiation from nearby Pop III stars can ablate the outer layers 
of the halo in supersonic flows and expose its interior to the IGM \citep{il04,il05,oet05,su06,su07,
wet08b,hus09,suh09,wet10}.  The dense core casts a cylindrical shadow that is eventually crushed 
down into the axis of the halo by thermal pressure forces in the surrounding ionized gas.  If metals 
from a SN collide with the ablative flow they can mix violently with it and cool it, perhaps causing it 
to fragment and collapse into new stars.  The metals can also envelop the partially exposed core, 
enriching and cooling it to greater depths than those found by \citet{cr08}.  The collision of SN 
ejecta with neighboring halos may therefore trigger new star formation, not just enrich the halo.

How metals mix with halos in such scenarios is also crucial to predicting elemental abundances in 
the second-generation stars that do form, some of which may still exist today as ancient, dim metal
poor stars in the Galactic halo \citep[e.g.,][]{Cayrel2004,Lai2008,Frebel10,nor13,cau13,bf15,ao15}. 
If the abundances found in extremely metal-poor stars can be reconciled to the nuclear 
yields of Pop III SNe they can place constraints on the masses of the first stars \citep{Iwamoto2005,
cooke11,caffau12,cm14,cpj15}.  So far, the fossil abundance record suggests that early chemical enrichment 
is due more to core-collapse (CC) than massive pair-instability (PI) SNe \citep{jet09b}, 
but see \citet{karl08,aoki14}.  Past studies have focused just on the yields, not the dilution factors 
with which metals are found in stars. Numerical simulations that resolve both mixing and cooling in
pristine gas are needed to reproduce these concentrations in specific stars, such as J031300
\citep{keller14,chen17a}. 

Here we examine the collision of ejecta from Pop III SNe with nearby halos as a potential channel 
of prompt second-generation star formation in the early universe with the \ZEUS\ and \CASTRO\ 
codes.  Our simulations include the evaporation of the halo by the star prior to the explosion and 
resolve mixing on spatial scales that are not currently practical in cosmological simulations.  In 
Section 2 we describe our halo evaporation and SN blast models, which were done with \ZEUS.  
Our mixing calculations with \CASTRO\ are discussed in Section 3. In Section 4 we examine how 
metals from CC and PI SNe mix with halos with a variety of masses and evaporation states.  We 
discuss the prospects for second-generation star formation after the collision and conclude in 
Section 5.  

\section{Numerical Method}

Our simulations are carried out in two stages.  First, the halo is photoionized by the external star 
with the \ZEUS\ cosmological radiation transport code.  The collision of ejecta from the explosion 
of this star with the ionized halo is then modeled with \CASTRO.  In \CASTRO, the SN ejecta is 
treated as a time-dependent, plane-wave inflow at the lower boundary of the simulation box 
whose that is extracted from 1D expanding grid models of the SN in \ZEUS.  We model mixing in 
two stages because \ZEUS\ has the radiation transport required to evaporate the halo and 
\CASTRO\ has the adaptive mesh refinement required to resolve the collision of the ejecta with 
the halo.

\subsection{\ZEUS}

\ZEUS\ is a massively-parallel radiation hydrodynamics code that evolves astrophysical fluid flows 
self-consistently with nonequilibrium nine-species primordial gas chemistry and multifrequency 
photon-conserving raytracing UV transport to evolve cosmological ionization fronts \citep[I-fronts;]
[]{wn06,wn08b,wn08a}.  The halo photoevaporation models in this study are identical in physics
to those in \citet{wet08b} and \citet{wet10}; here, we review the properties of the stars, the halos 
and the simulation box in \ZEUS.

\begin{deluxetable}{lcccc}
\tablecaption{Target Halos and SNe}
\tablehead{
\colhead{Model} &
\colhead{Halo mass} &
\colhead{R$_{200}$} &
\colhead{SN Type} 
\\
\colhead{} &
\colhead{($\Ms$)} &
\colhead{(pc)}    &
\colhead{}    
}
\startdata
	h1sn25    &  $6.9 \times 10^5$  &  249.6  & CC  \\
	h1sn200  &  $6.9 \times 10^5$  &  249.6  & PI    \\
	h2sn25    &  $2.1 \times 10^6$  &  320.1  & CC  \\
	h2sn200  &  $2.1 \times 10^6$  &  320.1  & PI    \\
	h3sn25    &  $1.2 \times 10^7$  &  483.1  & CC  \\
	h3sn200  &  $1.2 \times 10^7$  &  483.1  & PI       
\enddata
\tablecomments{R$_{200}$ = R($\rho = 200 \, \Omega_{b,0} \rho_{crit}$).}
\label{tbl:models}
\end{deluxetable}

\begin{figure}
	\plotone{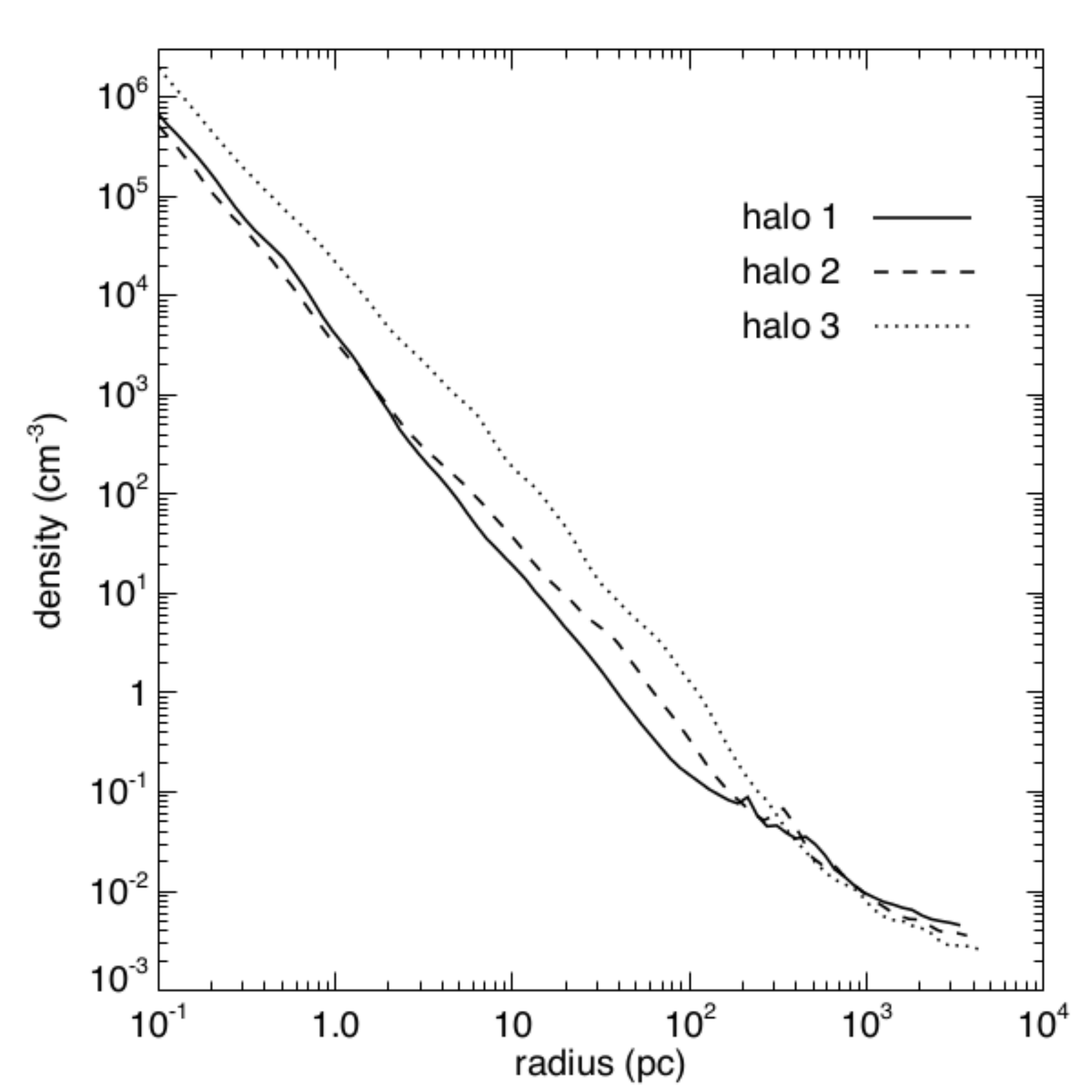}
	\caption{Spherically averaged density profiles for the 6.9 $\times$ 10$^5$ \Ms\ (halo 1), 2.1 $\times$ 
		10$^6$ \Ms\ (halo 2), and 1.2 $\times$ 10$^7$ \Ms\ (halo 3) models.} \vspace{0.1in}
	\label{fig:rho}
\end{figure} 

\begin{figure*}
\plotone{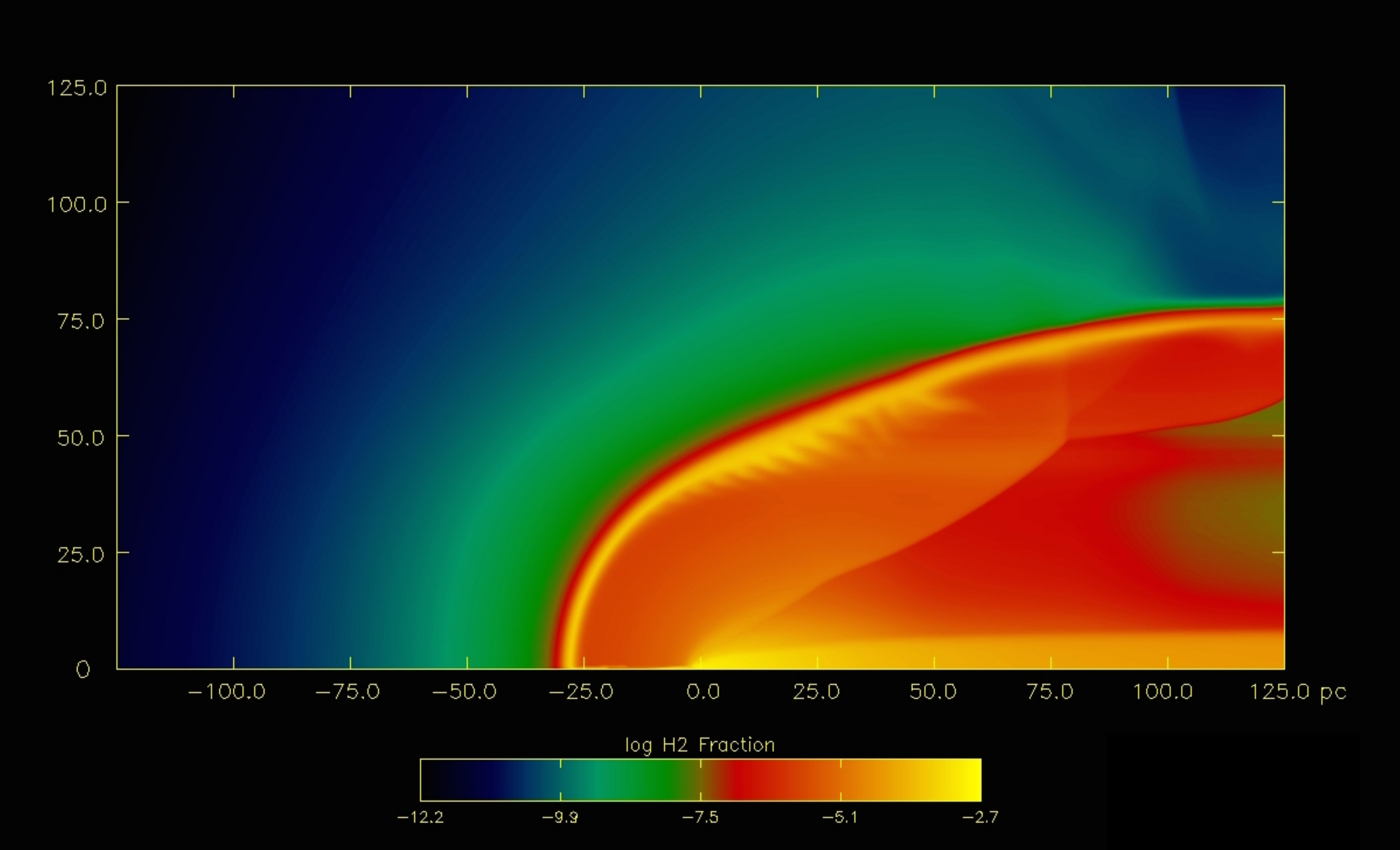}
\caption{H$_2$ mass fractions for the partially evaporated 6.9 $\times$ 10$^5$ \Ms\ halo that is
250 pc from the 25 \Ms\ star at the end of the life of the star. The horizontal and vertical axes are 
$z$ and $r$, respectively.} \vspace{0.1in}
\label{fig:halo}
\end{figure*} 

\subsubsection{Halo Models}

We consider 25 and 200 \Ms\ stars at a distance of 250 pc to 6.9 $\times$ 10$^5$, 2.1 $\times$ 
10$^6$, and 1.2 $\times$ 10$^7$ \Ms\ halos.  These halos bracket the range in mass in which 
Pop III stars form via H$_2$ cooling and are extracted from cosmological simulations performed 
with the Enzo code \citep{enzo}. \djw{We show spherically-averaged density profiles of all three 
halos prior to illumination in Figure~\ref{fig:rho}.  As seen from the central densities, they are at 
intermediate stages of collapse but have not yet formed stars.} The densities, gas energies, velocities and species mass 
fractions of these halos are spherically averaged and mapped onto a two-dimensional (2D) 
cylindrical coordinate grid in \ZEUS.  The ionizing UV fluxes and lifetimes of the two stars are 
taken from \citet{s02}.  Each halo is centered on the $z$-axis, with only its upper hemisphere 
residing on the grid. The grid has 1000 zones in $z$ and 500 zones in $r$ for a spatial resolution 
of 0.25 pc.  The boundaries are at -125 pc and 125 pc in $z$ and 0.01 pc and 125 pc in $r$.  
Reflecting boundary conditions (BCs) are applied to the inner $r$ boundary and outflow BCs are 
set on the other three boundaries.  All simulations were performed at redshift $z =$ 20.

\subsubsection{Photoevaporated Halos}

We show H$_2$ mass fractions for the partially evaporated halo when the star dies in Figure 
\ref{fig:halo} \citep[the process of halo photoevaporation is discussed in greater detail in][]
{wet08b,wet10}.  The D-type I-front has a cometary appearance because it preferentially 
advances in the more stratified densities above and below the midplane of the halo.  A 
prominent arc of H$_2$ is visible in the outer layers of the I-front. H$_2$ forms there because 
the hard UV spectrum of the star broadens the I-front, creating temperatures of a few thousand 
K and ionization fractions of $\sim$ 0.1 in its outer layers that catalyze the rapid formation of 
H$_2$ via the H$^{-}$ channel \citep{rgs01}.  The H$_2$ in the outer layers of the I-front 
partially shields the core from Lyman-Werner (LW) photons from the star, as shown by the 
large H$_2$ mass fractions there.  The core also shields gas behind it from LW flux.  

The serrated structures midway up the arc of the front are remnants of dynamical instabilities that 
are driven by H$_2$ cooling in the outer layers of the front.  They are more prominent further out 
along the arc because of the larger angles of incidence between the photons and the front there
\citep{rjr02}. These structures are of interest because they enhance mixing between outflows from 
the halo and the SN ejecta.  Supersonic flows are also driven back towards the star as the outer 
layers of the halo are blown off by the radiation front.  These flows collide with and mix with ejecta 
from the SN from the left, as discussed later. 

After the star dies the halo and H II region are evolved in \ZEUS\ for the additional time required for 
metals from the SN to reach the left boundary of the grid in \CASTRO.  The pressure of the ionized 
gas continues to crush the shadow cast by the halo downward toward its axis after the explosion
and the supersonic backflows keep expanding towards the ejecta.  The ionized gas also begins to 
recombine out of equilibrium, \djw{cooling more quickly than it recombines.}  Following these three
processes for the time it takes debris from the SN to reach the boundary of the \CASTRO\ box 
ensures that the metals encounter a more realistic circumhalo structure at the time of collision. The 
time required for the ejecta to cross this distance is taken from the 1D \ZEUS\ SN models  
described below.  The halos are illuminated by the 25 \Ms\ star for its main-sequence lifetime of 
6.46 Myr, after which radiation is switched off and the H II region is allowed to cool and recombine 
for an additional 2.16 Myr. For the 200 \Ms\ star these two times are 2.2 Myr and 0.872 Myr, 
respectively. 

\begin{figure*}
\plottwo{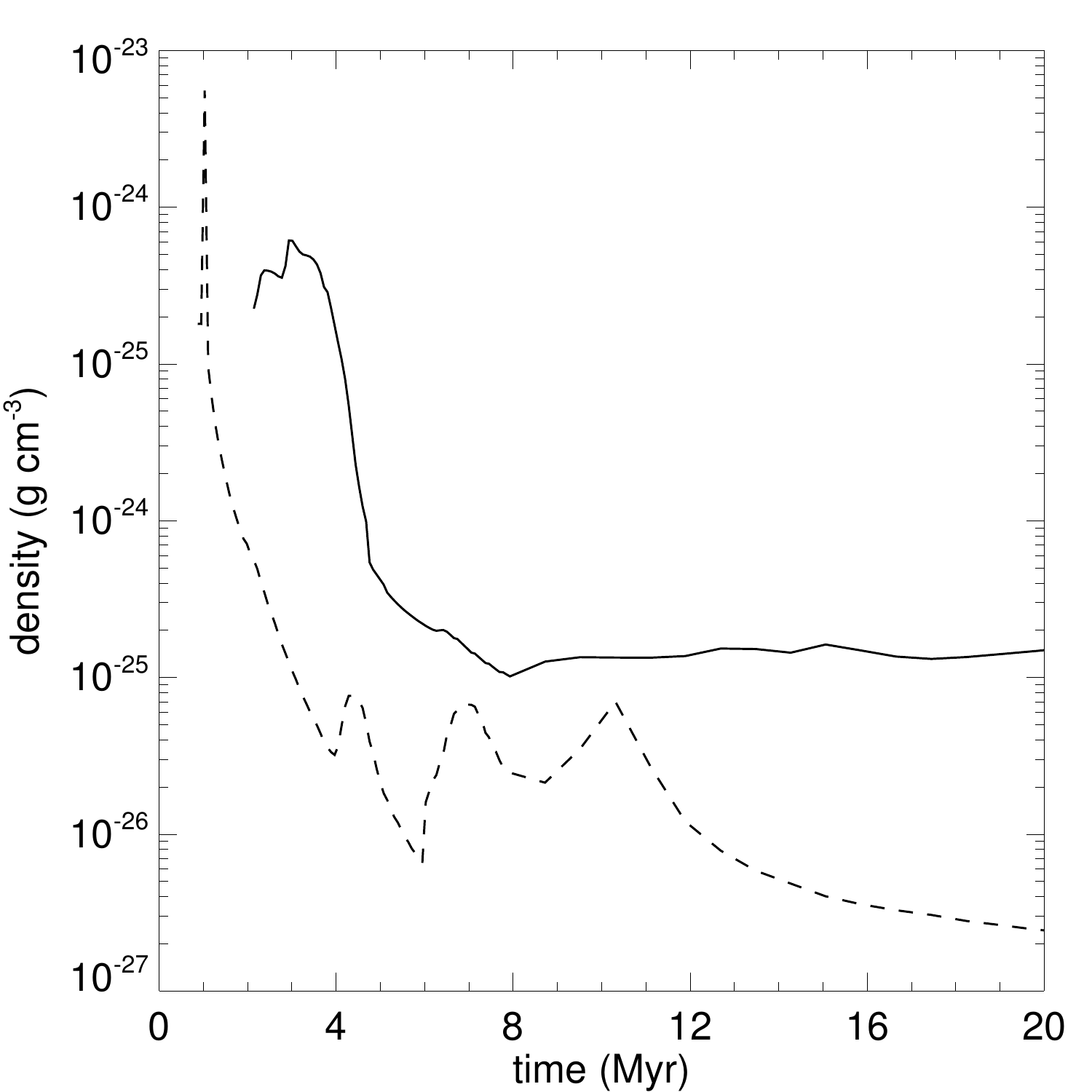} {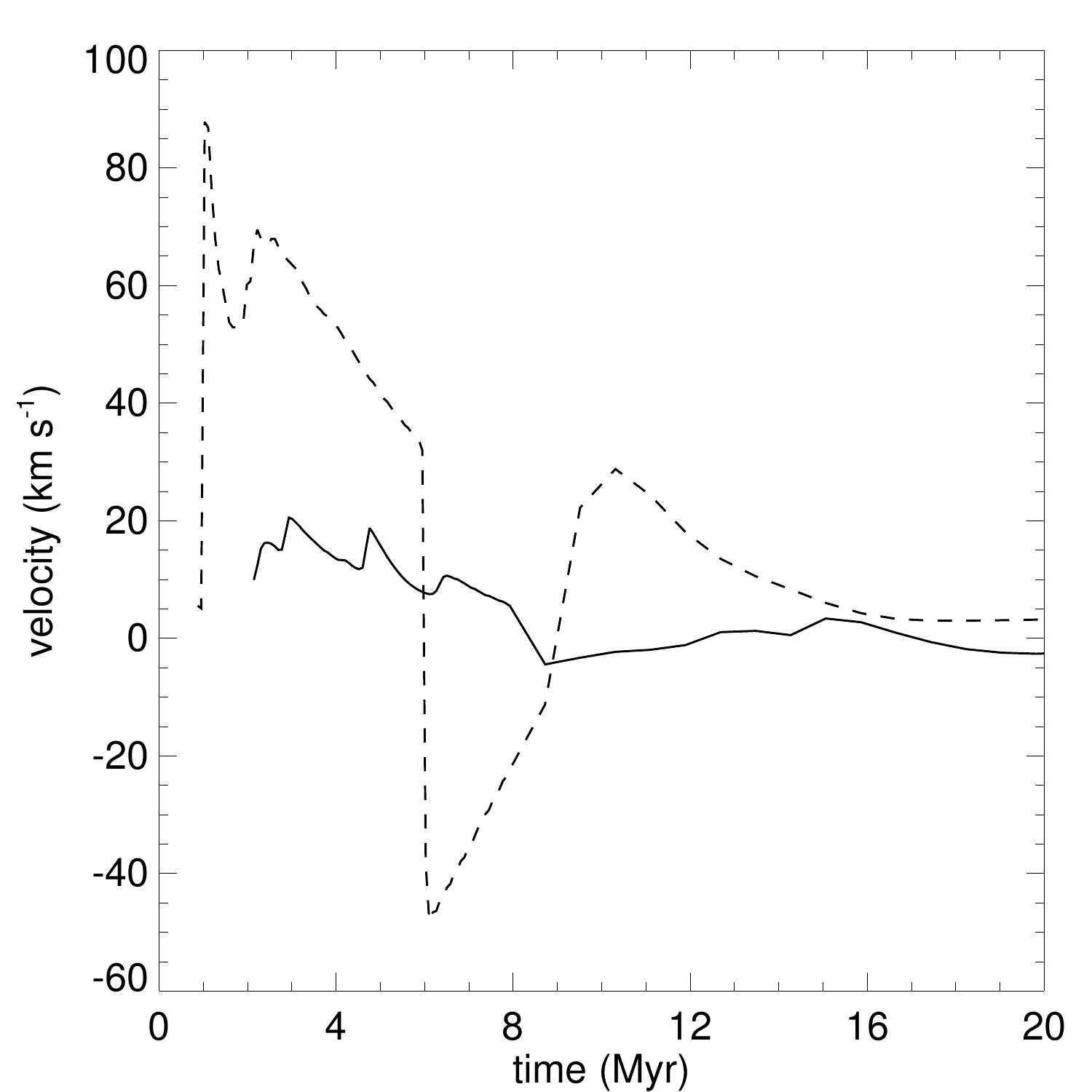}
\caption{Time dependent boundary conditions representing the flow of ejecta from a Pop III SN 
past a fixed observer position at 125 pc, derived from 1D expanding grid explosion models done 
with \ZEUS.  Both explosions occur 250 pc from the halo. Left panel: densities. Dashed line: the 
200 \Ms\ PI SN.  Solid line: the 25 \Ms\ CC SN.  Right panel: velocities.  Dashed line: the 200 
\Ms\ PI SN. Solid line: the 25 \Ms\ CC SN.} \vspace{0.1in}
\label{fig:tdbc}
\end{figure*} 

\subsection{\CASTRO}

The halo and surrounding H II region are mapped onto a 2D cartesian grid in \CASTRO\ with 
a bilinear interpolation scheme. Since only the upper half of the halo is modeled in \ZEUS, 
its lower half is assumed to be the mirror image of its upper half.  \CASTRO\ is a 
multidimensional adaptive mesh refinement (AMR) radiation hydrodynamics code \citep{
almgren2010,zhang11} with an unsplit piecewise parabolic hydro scheme \citep{wc84} and 
an ideal gas equation of state with $\gamma$ = 5/3.  We advect three species (hydrogen, 
helium, and metals) and use a simple cooling function from \citet{Gne06} but neglect the 
detailed primordial gas chemistry and cooling and metal line cooling.  The mass fractions of 
the five H species in the \ZEUS\ models are summed to obtain the single H species in 
\CASTRO\ and the three He species are summed to obtain the one He species. The coarse 
grid is 250 pc in $x$ and $y$ with 1024$^2$ zones.  Up to four levels of refinement are 
allowed for a factor of 16 (2$^4$) higher spatial resolution.  Refinements are performed on 
gradients in density and velocity.   

Four nested grids are centered on the halo for a maximum resolution of 0.015 pc, or 3000 AU.  
Inflow BCs are set on the lower $y$ boundary, where SN ejecta flow onto the grid as described 
below, and outflow BCs are set on the other three boundaries so metals and ionized gas can exit 
the grid.  The two explosions, the 25 \Ms\ CC SN and the 200 \Ms\ PI SN, have energies of 1 
B and 42 B, respectively (1 B $= 10^{51}$ erg).  The halo and metals are evolved for 20 Myr. We 
summarize our models in Table \ref{tbl:models}.  These 2D simulations required 480,000 CPU 
hours on the Cray XC-30 machines at the National Energy Research Scientific Computing Center 
(NERSC) and the Center for Computational Astrophysics (CfCA) at the National Astronomical 
Observatory of Japan (NAOJ).

\subsubsection{SN Inflow Model}

We treat the SN ejecta as a time-dependent planar inflow on the lower $y$ boundary. Gas densities, 
energies, velocities and mass fractions for all three species are updated on the boundary every time 
step during the simulation.  These time-dependent flows are derived from 1D explosion models of 
the 25 and 200 \Ms\ SNe in \ZEUS. In these 1D models, the SN is evolved in its own halo as in 
\citet{wet08a}, but only out to radii corresponding to the position of the lower boundary of the box in 
\CASTRO.  These simulations are done in two steps. First, the H II region of the star in its own halo 
is created as described in \citet{wet08a}, \citet{anna15} and \citet{anna17} with ionizing photon rates
again taken from \citet{s02} for consistency.  The host halos of the 25 and 200 \Ms\ stars are taken 
to be identical to the 6.9 $\times$ 10$^5$ and 2.1 $\times$ 10$^6$ \Ms\ neighbor halos, respectively.  
This choice of host halos is arbitrary, but reasonable.  Both stars ionize their own halos within a few 
hundred kyr, with supersonic flows from the core plowing most of the baryons into a dense shell with 
a radius of 100 - 200 pc by the time the star dies.  At this point the densities at the center of the halo 
are low, $\sim$ 0.1 - 1 cm$^{-3}$ \citep[e.g., Fig. 3 of][]{wan04}, so both stars explode in diffuse 
media.

\begin{figure}
	\plotone{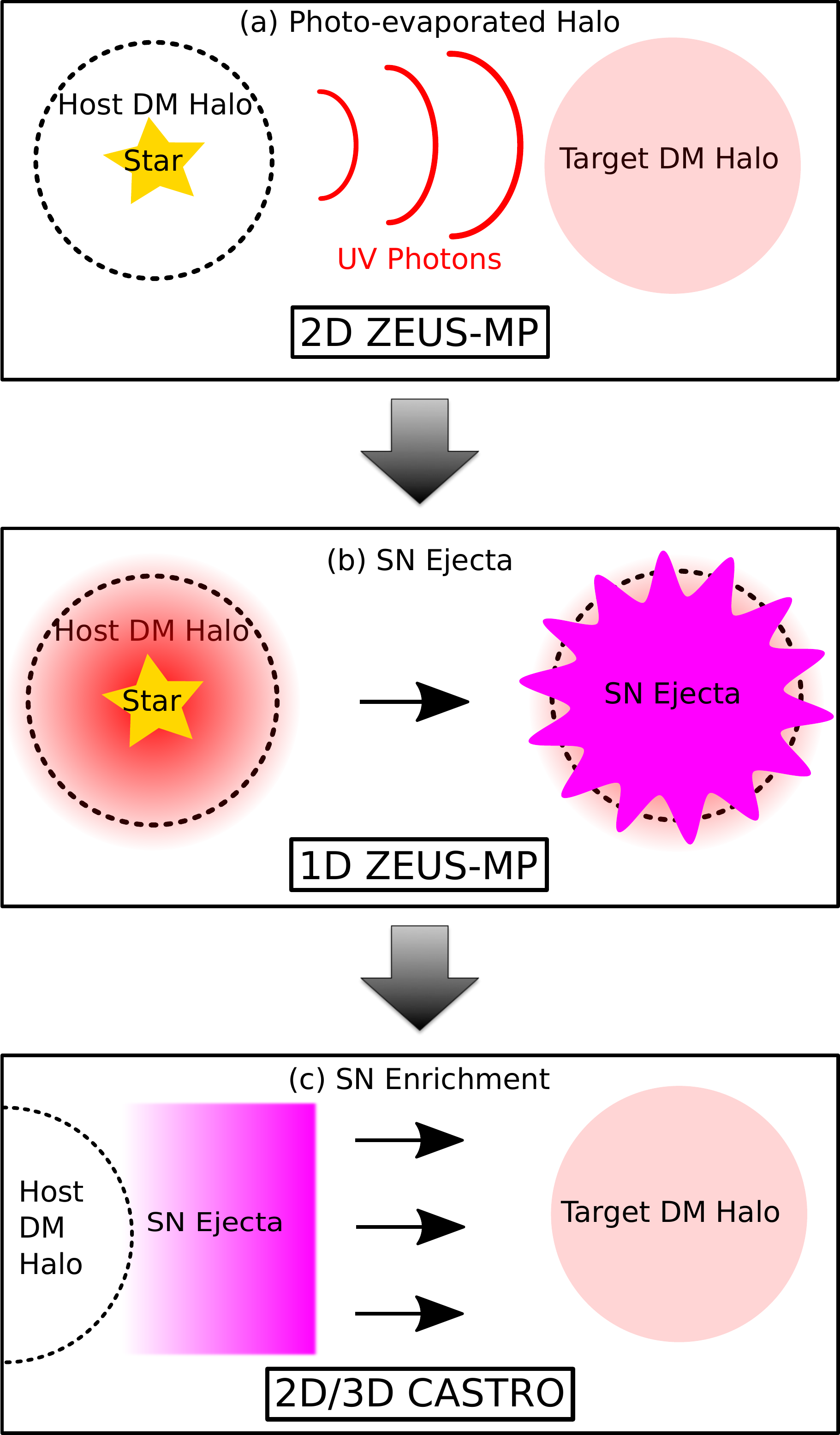}
	\caption{\ken{Simulation flowchart.  Our numerical setup can be divided into three 
			steps shown in Panels (a-c).  We first use 2D \ZEUS\ to simulate UV feedback to the target
			halo until the star dies as a SN, then simulate the propagation of SN ejecta 
			in 1D \ZEUS\ before they reach the boundary of \CASTRO\ box.  Finally, we use the initial and boundary 
			conditions from two previous steps to simulate the cosmological mixing in multi-dimension with \CASTRO.}} \vspace{0.1in}
	\lFig{method}
\end{figure}

\begin{figure*}
\plotone{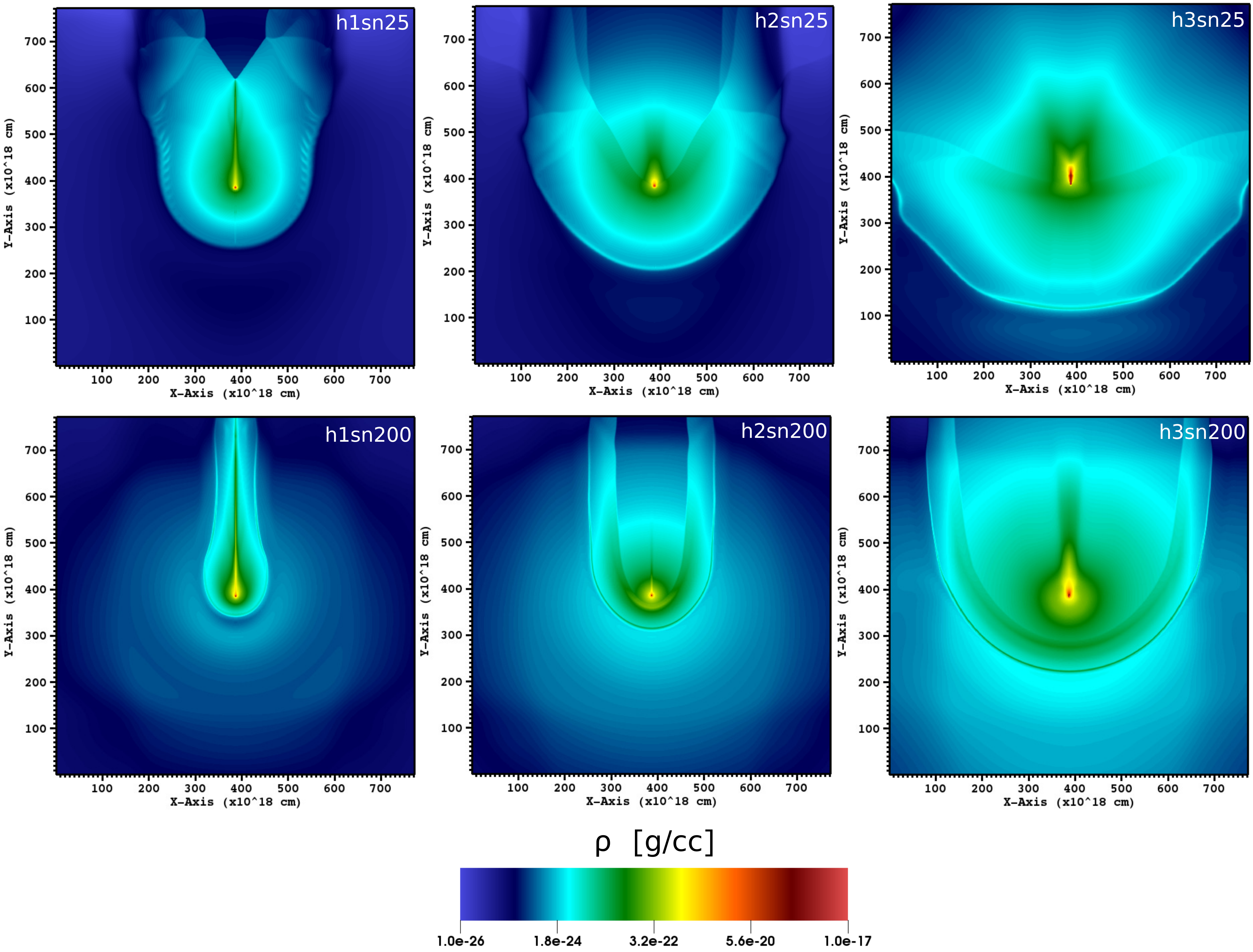}
\caption{Density images of the photoevaporated halos after the death of the star, when ejecta 
from the explosion have reached the lower boundary of the grid in \CASTRO.  The degree to 
which they are ionized depends on their mass and stellar luminosity. The envelope of the halo 
in h1sn200 has been mostly stripped away while the envelope of h3sn25 is only slightly 
affected.} \vspace{0.1in}
\lFig{halo_den}
\end{figure*} 

We then explode the star in the center of its H II region on a 1D expanding grid in \ZEUS\ 
as explained in \citet{wet08a} \citep[see also][]{mw12}.  The blast wave is initialized as a 
free expansion with density and velocity profiles from \citet{tm99}, which are parametrized 
by ejecta mass, $M_{\mathrm{ej}}$, explosion energy, $E_{\mathrm{ej}}$, and peak ejecta 
velocity, $v_{\mathrm{max}}$.  We take $M_{\mathrm{ej}} =$ 20 \Ms\ and $E_{\mathrm{ej}} 
=$ 1 B for the 25 \Ms\ SN (assuming the formation of a 5 \Ms\ black hole during the core
collapse) and $M_{\mathrm{ej}}=$ 200 \Ms\ and $E_{\mathrm{ej}} =$ 42 B for the 200 \Ms\ 
star \citep[no compact remnant;][]{hw02}. In both cases $v_{\mathrm{max}} =$ 3.0 $\times$ 
10$^9$ cm s$^{-1}$.  The profiles were initialized on a 1D spherical grid with 250 uniform 
zones and an outer boundary at 4.626 $\times$ 10$^{15}$ cm. These models are otherwise 
identical in physics to those in \citet{wet08a} and \citet{mw12}, in which the importance of 
inverse Compton cooling to the evolution of Pop III SN remnants in the primordial universe 
is reviewed in detail \citep[see also][]{ky05}. 

The SN remains essentially a free expansion until it has swept up about its own mass in 
ambient gas at radii of 10 - 20 pc.  At this point a reverse shock forms and is driven back 
toward the center of the halo in the frame of the shock, reverberating back and forth 
through the ejecta over time in some runs.  When the ejecta crash into the dense shell of 
the relic H II region the collision can drive another reverse shock into the interior of the 
halo. To an observer at a fixed radius (such as at the lower boundary of the \CASTRO\ 
box), passing flows would therefore be highly time dependent, with strong forward flows 
at some times and backflow at others. 

We show velocities and densities at 125 pc, the position of the lower boundary in \CASTRO, 
for the 25 \Ms\ CC SN and 200 \Ms\ PI SN in Figure~\ref{fig:tdbc}. They are extracted from 
the 1D \ZEUS\ run over seven or eight decades in time at uniform intervals in log time. The 
passage of the shock and the rarefied regions in its wake are visible as the peaks and valleys 
in density.  The peak happens earlier in the PI SN flow because it is more energetic and 
reaches the outer regions of the halo sooner.  Both forward and reverse shocks are visible in 
the velocity profiles, with the latter causing negative velocities in both flows.  To compute the 
SN ejecta entering the \CASTRO\ grid, the densities and velocities shown in 
Figure~\ref{fig:tdbc} (and energy densities and mass fractions) are stored in a table as a 
function of time.  At a given time step in \CASTRO\ the inflow boundary conditions are 
computed with a linear interpolation of the logarithm of the table entries vs. log time.  

\ken{Our simulations using two different codes with different dimensions and physics are 
		conceptually complicated. Therefore, we use a cartoon to summarize the entire 
		numerical setup in Figure~\ref{fig:method} to make it clearer. }

\section{Halo Enrichment}

\begin{figure*}
\plotone{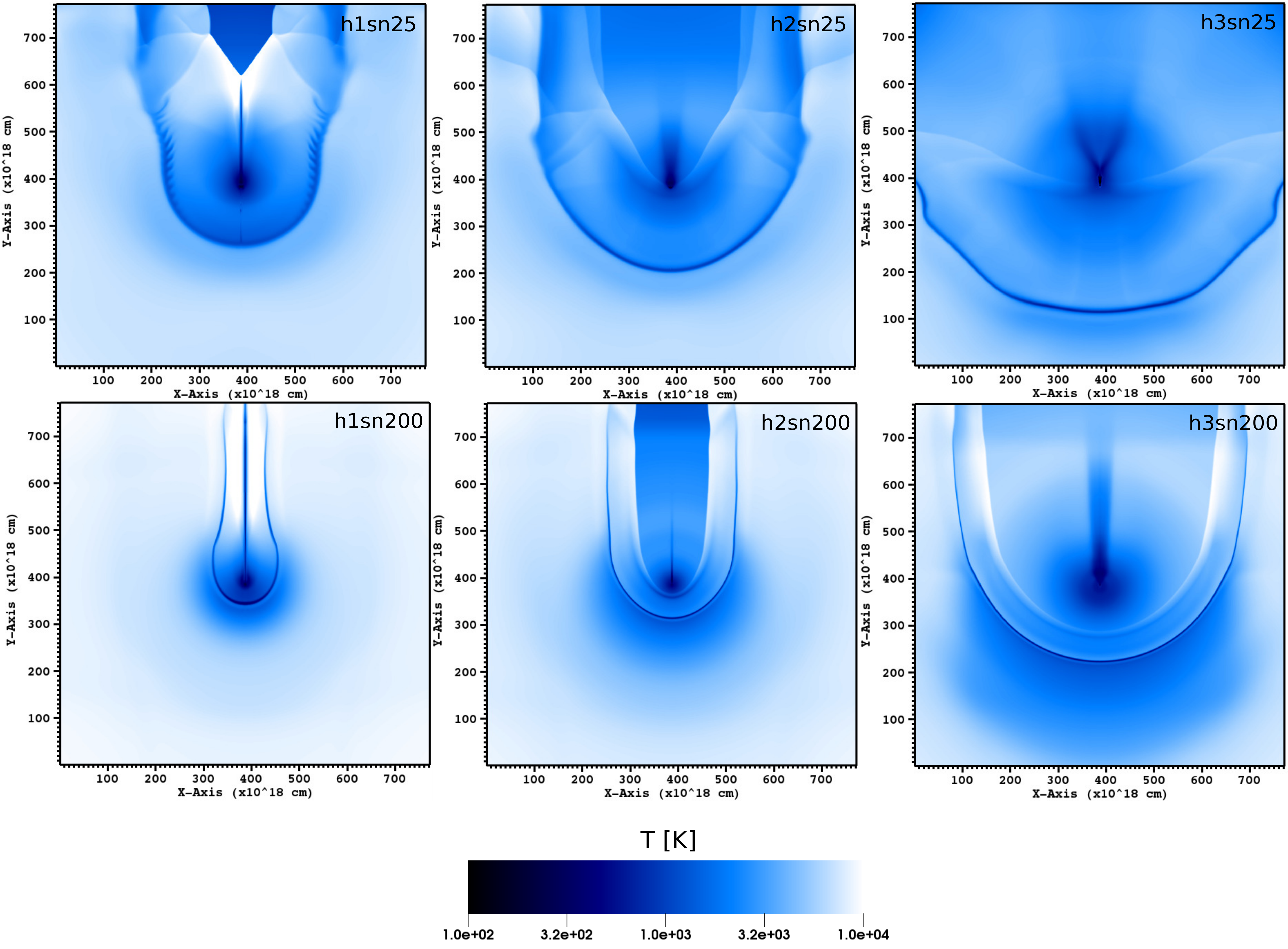}
\caption{Temperature images of the photoevaporated halos after the death of the star, when ejecta 
from the explosion have reached the lower boundary of the grid in \CASTRO.  The partially ionized
and cooled gas surrounding the halos has a temperature a few thousand K, but their centers are 
only a few hundred K due because they are shielded from stellar radiation so H$_2$ cooling can 
occur.} \vspace{0.1in}
\lFig{halo_temp}
\end{figure*} 

\subsection{Halo Structure at Collision}

We show density and temperature images of the partially evaporated halos and their ionized 
envelopes at the time the SN enters the lower boundary of the grid in \CASTRO\ in Figures
\ref{fig:halo_den} and \ref{fig:halo_temp}.  After the star dies the ionized gas crushes the shadow 
cast by the halo down towards its axis, even as it cools and recombines.  The solid angle cast by 
the halo varies with its mass, proximity to the star, and the luminosity of the star.  More massive 
halos and less luminous stars cast the broadest shadows while less massive halos and more 
luminous stars create narrow cocoons with an almost cometary appearance.  The dense ridges 
of gas created by I-front instabilities in Figure~\ref{fig:halo} are visible in the h1sn25 profile but do 
not appear in the other halos. In general, more massive stars with higher fluxes create less H$_2$ 
in the outer layers of the I-front so there is less cooling and instability growth.  I-fronts from any 
given star are also less likely to form instabilities in more massive halos because it is more difficult 
for them to develop in higher densities.  

By this time, regions of the halo and its now partially ionized envelope have cooled to $\sim$ 150 K
because of rapid H$_2$ formation after the death of the star. In the absence of UV flux the ionized 
gas begins to recombine out of equilibrium, cooling more quickly than it becomes neutral.  Denser
regions of the gas fall to temperatures and ionized fractions of a few thousand K and $\sim$ 0.1 
first, triggering rapid H$_2$ formation via the H$^-$ channel.  This process is what cools gas in the 
ionized envelope of the halo.  Temperatures in the core and its shadow fall over time even though 
they were never ionized because they are no longer exposed to LW flux from the star, so H$_2$ 
rapidly forms in the high densities there and cools the gas.  The cold gas in the thin layers tracing 
the remnant of the I-front is due to H$_2$ that formed there while the star was alive.  It is coldest 
because the H$_2$ is located in the dense gas plowed up by the front.  Low gas temperatures in 
Figure~\ref{fig:halo_temp} track regions of higher density in Figure~\ref{fig:halo_den} because H$
_2$ formation is a two-body process.  

\djw{We include H$_2$ cooling in the relic H II region because it can enhance mixing during collision 
with SN ejecta because it condenses the gas to higher densities.  Note that had HD chemistry 
been included in our models it would have cooled the gas even further, down to the CMB 
temperature ($\sim$ 57 K at $z =$ 20), because the non-symmetric HD molecule can radiate via 
dipole transitions while H$_2$ can only emit via quadrupole transitions \citep[e.g.,][]{ga08}. Lower 
temperatures might have resulted in even higher densities and more efficient mixing. HD cooling 
is also known to lower the mass scales of fragmentation in relic H II regions \citep[see][]{yet07,
yoh07}, which leads to the formation of Pop III stars that are somewhat lower in mass than those 
forming from H$_2$ cooling alone.}

\subsection{Collision of the SN with the Halo}

\begin{figure*}
\epsscale{1.9}
\plottwo{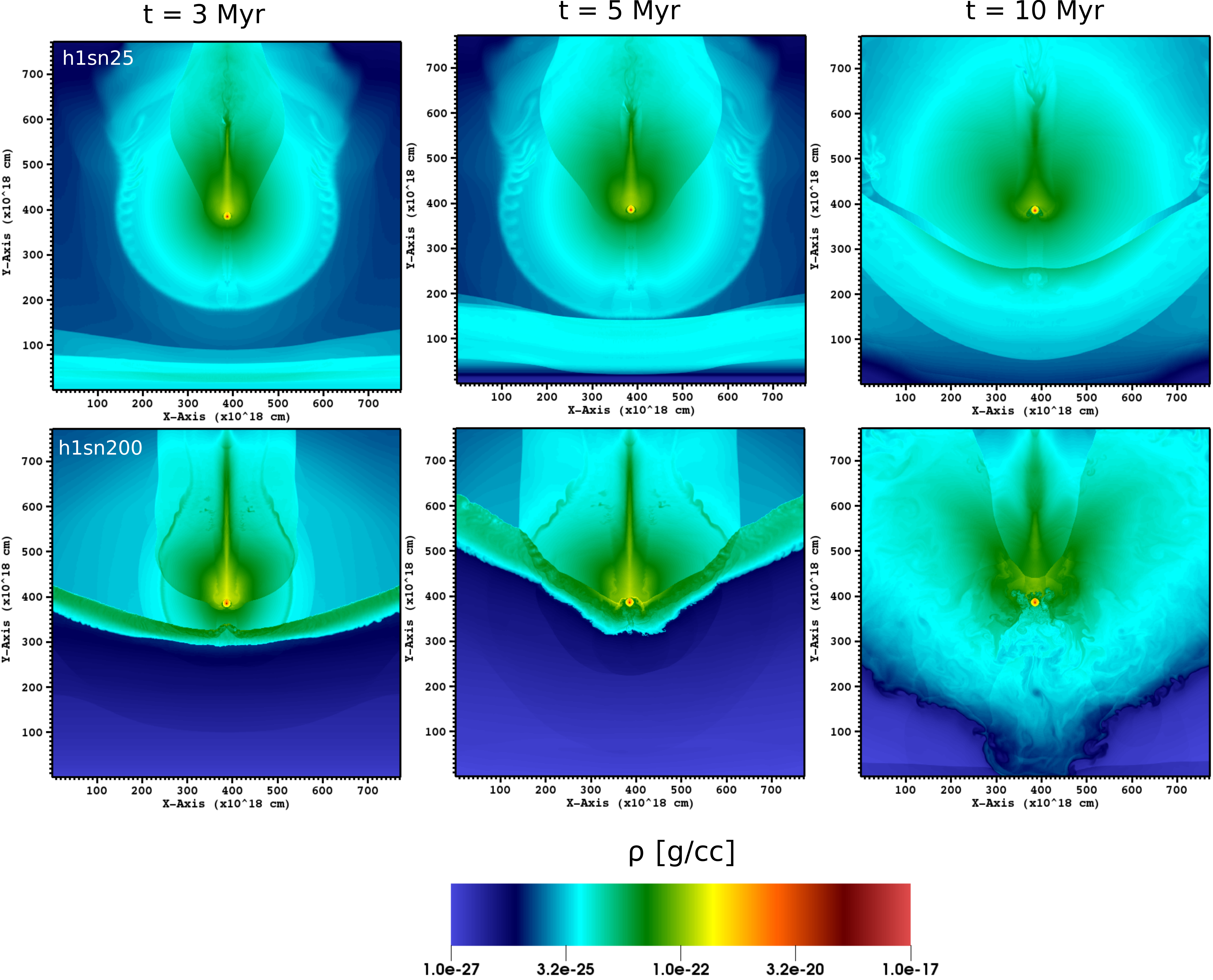}{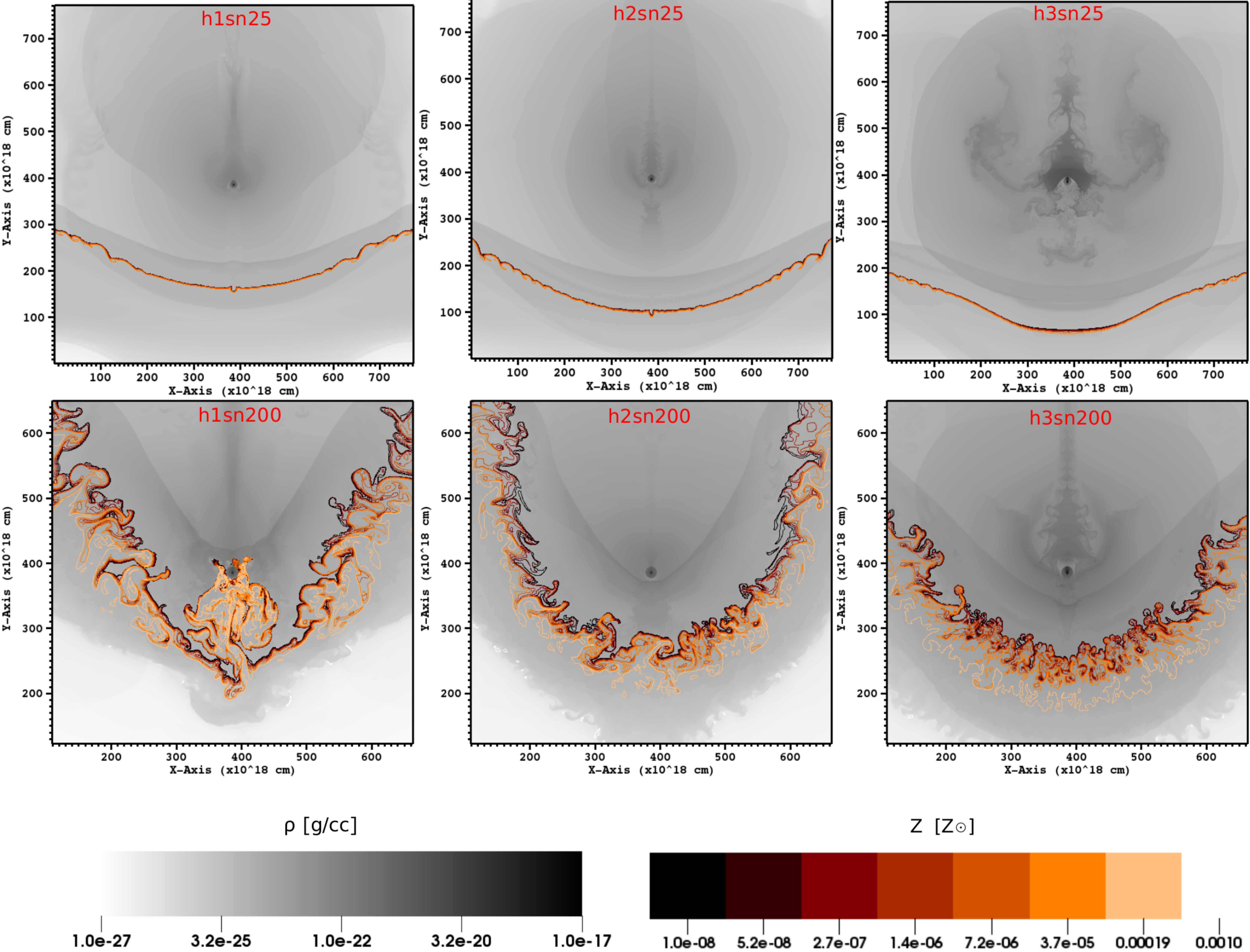}
\caption{Upper panels: collision of the SN with the halo at 3, 5 and 10 Myr. In the 25 \Ms\ runs 
the shock stalls 50 pc from the center of the halo but reaches and disrupts the center in the 200 
\Ms\ models.  Bottom panels:  distribution of metals in the halo at 10 Myr.  Metals are halted 70 
- 100 pc from the center of the halo in the 25 \Ms\ runs because of lower UV fluxes and 
explosion energies, so Pop II stars are less likely to form in the core but may occur in an arc 
around it.  The 200 \Ms\ SN has a more violent impact with the halo and more prominent fluid 
instabilities appear.   Mixing enables metals to reach the core of the halo in the h1sn200 run.
 } \vspace{0.1in}
\lFig{halo_evo}
\end{figure*} 

\begin{figure}
\plotone{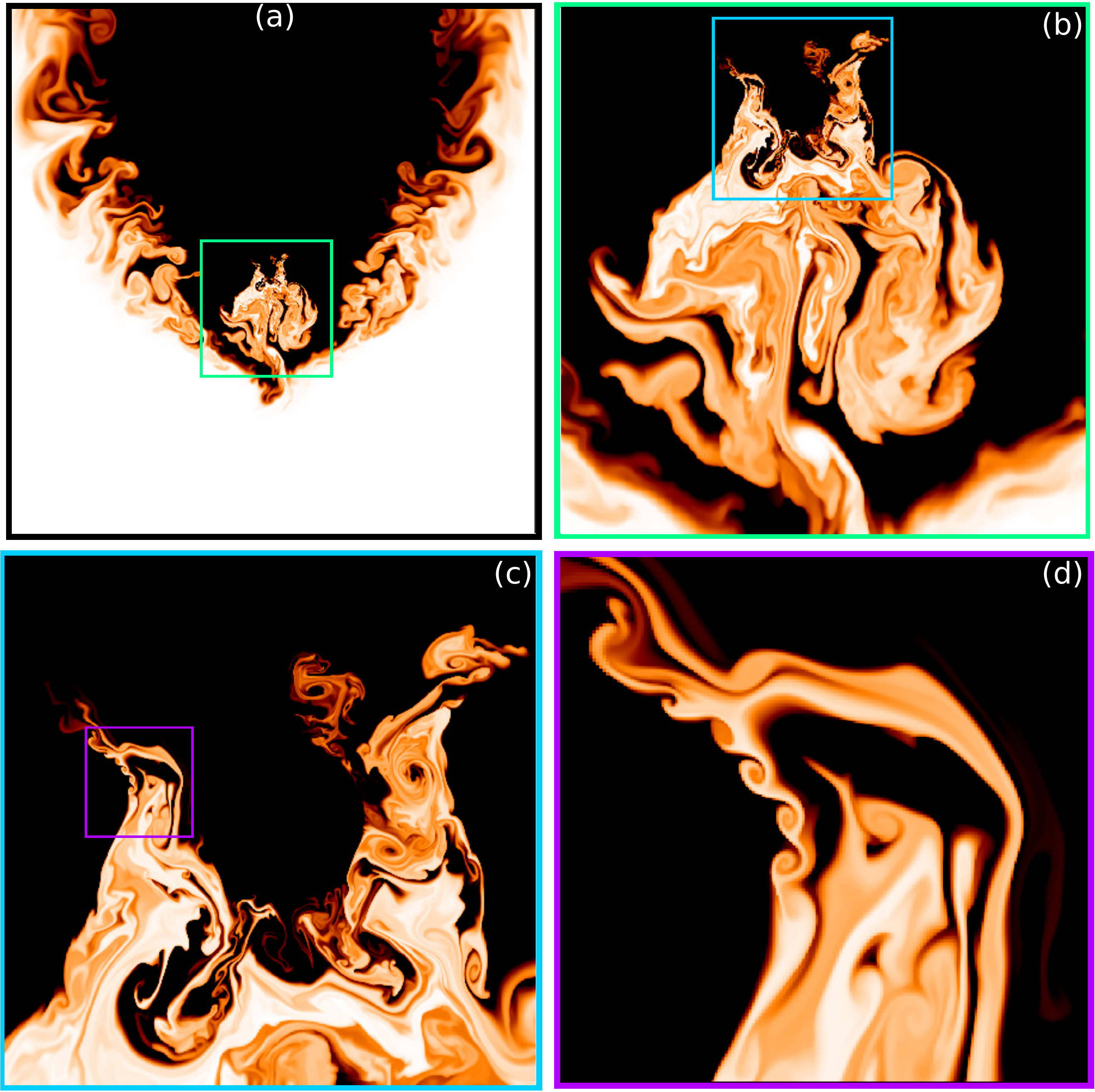}
\caption{Zoom in of mixing regions of h1sn200 at late times.  Panels a, b, c and d are 250, 80, 
20 and 4 pc across, respectively.  Mixing begins with Rayleigh-Taylor instabilities that develop
into Kelvin-Helmholz shear instabilities that later cause the flow to become highly turbulent.} 
\lFig{halo_zoom}
\end{figure} 

We show the collision of the SN with the halo in the h1sn25 and h1sn200 models in the upper 
panels of Figure~\ref{fig:halo_evo}.  Ejecta enter the lower boundary in \CASTRO\ at $\sim$ 20 
km s$^{-1}$ for the CC SN and $\sim$ 30 km s$^{-1}$ for the PI SN. As the shock approaches 
the halo it decelerates as it collides with the photoevaporative backflow and then climbs up the 
density gradient of the halo, advancing to within $\sim$ 100 pc of its core by 3 - 5 Myr.  In the 
h1sn25 run the shock stalls 50 pc from the center of the halo.  The h1sn200 shock is more 
energetic and gets to within 5 pc of the center. In both runs the stripping of the outer layers of 
the halo by ionizing UV photons from the star allows metals to reach greater depths in the halo 
than would otherwise have been possible.  The remnants of the I-front instabilities that are 
visible in the density image of the h1sn25 model at 5 Myr develop into an extended turbulent 
layer as the SN shock overruns the system.  This structure dominates the dynamics of the 
entire envelope at 10 Myr and results in very efficient mixing.

If the shock is strong enough and the densities it encounters are not too high, a reverse shock 
s and detaches from the forward shock as it plows up more gas.  The contact discontinuity 
separating the two shocks crumples into Rayleigh-Taylor (RT) instabilities that in turn drive 
Kelvin-Helmholz (KH) shear instabilities, and the metals begin to mix with pristine gas in the 
halo.  If the shock is weaker or less of the halo is evaporated by the star, the RT instabilities, if 
they form, are milder and less mixing occurs.  Both cases are shown in the lower panels of 
Figure~\ref{fig:halo_evo}.  The stronger PI SN shock drives more mixing than the CC SN in a 
given halo.  Mixing is weaker in more massive halos for a given SN because they lose less gas 
to photoevaporation prior to the explosion and present more mass to the shock.  

The most mixing occurs in h1sn200, in which metals from the SN reach the core of the halo. 
In this model only metal-enriched Pop II stars would form, in the layers facing the explosion 
and in its center. Metals do not reach the core in the h2sn200 and h3sn200 runs so both Pop 
III and Pop II stars may not form in these models.  In contrast, Pop II star formation, if it occurs, 
would only happen in a narrow arc 50 - 100 pc in front of the center of the halo in the sn25 
runs.  The top panel of Fig.~\ref{fig:halo_evo} illustrates that this is the only region that has
sufficient metal enrichment.  One caveat is the relative timescales of Pop II and Pop III star 
formation in the h2sn200 and h3sn200 models. If Pop II stars form more quickly in regions 
with metals than Pop III stars form in the core, LW and H$^-$ photodetaching photons from 
the Pop II stars could destroy H$_2$ in the core, delaying cooling and collapse.  Ionizing 
photons could then later mechanically disrupt the core and prevent its collapse altogether.  
Improved simulations with metal and dust cooling that are able to resolve fragmentation and
collapse will determine the relative timescales of Pop II and III star formation in future models.

On the other hand, if a Pop III star forms in the core of the halo before its outer layers cool, 
fragment and collapse to Pop II stars they may never get the chance to form because UV 
from the Pop III star breaks out of the halo and completely ionizes it on timescales of a few 
hundred kyr. This order of events may apply particularly to some of the sn25 models in which 
metals do not come within 100 pc of the core because of weaker UV fluxes and explosion 
energies.  Our models therefore suggest that chemically enriched star formation in halos in 
close proximity to SNe should not be taken as a given, in contrast to what is usually assumed 
in semianalytical models and large-scale numerical simulations, since Pop III stars may form 
in them rather than Pop II stars in some cases.

\subsection{Distribution of Metals in the Halo}

\begin{figure}
\plotone{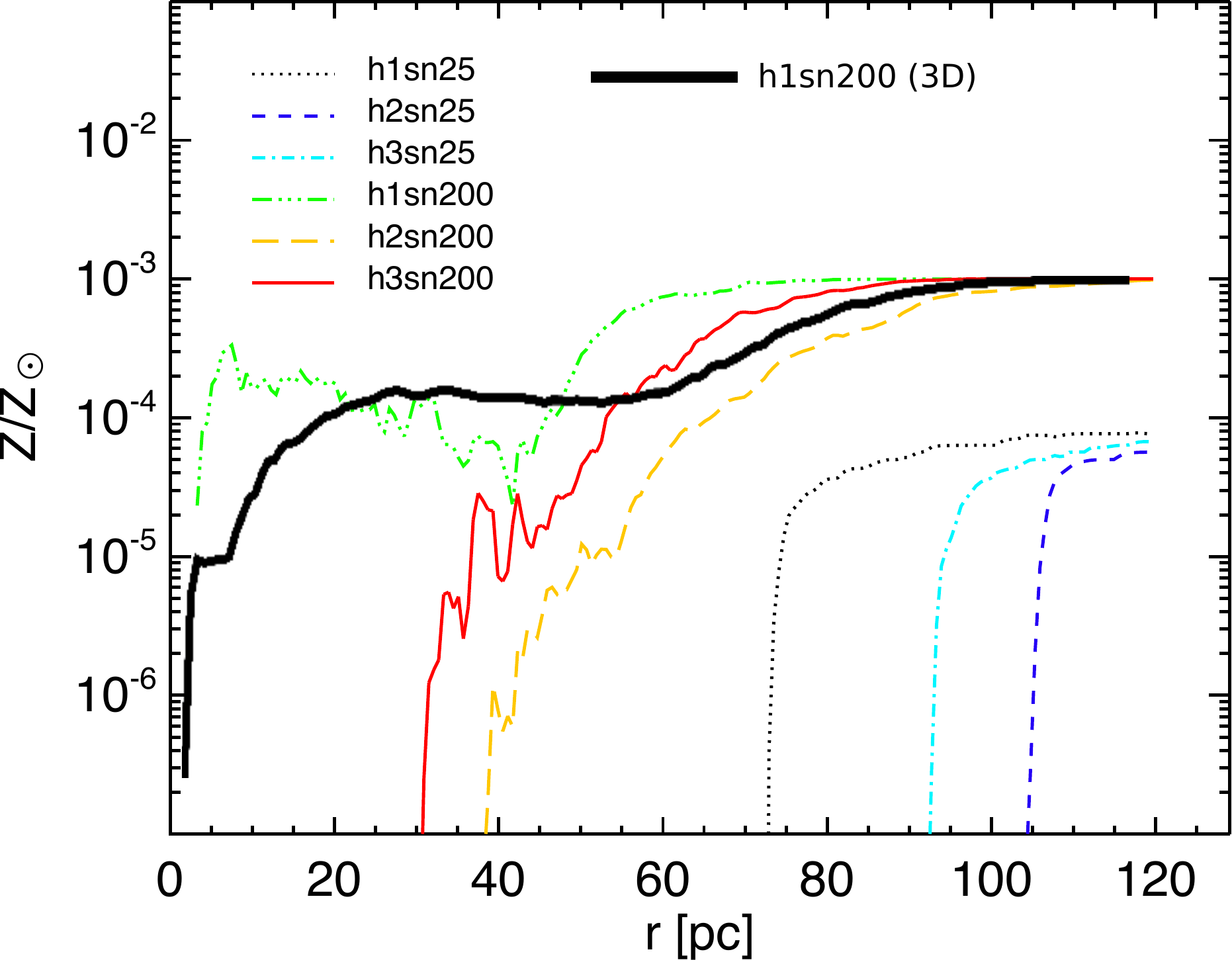}
\caption{Spherically-averaged metallicity profiles for the six 2D runs and the 3D h2sn200 
run at 10 Myr.  Metals from the 25 \Ms\ CC SN come no closer than 70 - 100 pc to the core 
in all three halos but come much closer to the core in the 200 \Ms\ runs, even reaching it in
the h1sn200 run. In this run a Pop II star could form at the center of the halo.  In the other 
five runs both Pop III and Pop II stars can form.  More metals reach the center of the halo 
in the h1sn200 model in 2D than in 3D.} \vspace{0.1in}
\lFig{met1}
\end{figure} 

We show mixing due to RT and KH instabilities on small scales in the h1sn200 model in Figure
\ref{fig:halo_zoom}.  Here, a fraction of the range of spatial scales on which mixing occurs is 
shown, with intricate structures with large surface areas that efficiently fold metals in with H and
He gas on scales of 250, 80, 20 and 4 pc. To quantify mixing, radially-averaged metallicities are 
plotted as a function of distance from the center of the halo for all six runs in Figure~\ref{fig:met1}. 
Metals from the three PI SNe reach greater depths in the core and enrich it to higher metallicites 
than the three CC SNe because of their higher energies and much larger metal yields, $\sim$ 50 
\Ms\ vs. a few \Ms.  The regions that are enriched by CC SNe on average only have metallicities 
of 10$^{-4}$ - 10$^{-5}$ \Zs\ and will only form Pop II stars if dust is present \citep[e.g.,][]{schn06,
kgc12}.  It is therefore again not a given that Pop II stars will form, although some gas in the halo 
will be at metallicities above the spherical average and may form such stars.  The disparity in 
mixing between PI and CC SNe is striking, with metallicities tapering off slowly at the center in 
the sn200 models but falling sharply at 70 - 100 pc in the sn25 runs.  Even in the core of the halo 
in the h1sn200 model the metallicity is 10$^{-3}$ - 10$^{-4}$ \Zs, enough to trigger the formation
of a Pop II star. \ken{We define the mixing efficiency, $\epsilon(z) = M_{\rm met}(z)/M_{\rm halo}$, $M_{\rm met}(z)$
is the mass of metal-enriched gas at a minimum metallicity $z$, $M_{\rm halo}$ is the mass of photoevaporated halo.
Depending on the mass of halo and its stricture, $\epsilon$ of a 25\Ms\  CC SN is about  $2-5 \times10^{-3}$ 
at $z = 10^{-6}\Zs$. For the same metallicity, $\epsilon$ of a 200\Ms\ PI SN can be $3-6 \times10^{-2}$  which
is about ten times bigger than CC SNe. We list $\epsilon(z)$ values  for each model in 
Table \ref{tbl:met}.  }

\begin{deluxetable}{lcccc}
	\tablecaption{Metallicity and its enriched gas fraction $\epsilon(z)$}
	\tablehead{
		\colhead{Model} &
		\colhead{$10^{-3}\Zs$}   &
		\colhead{$10^{-4}\Zs$ }  &
		\colhead{$10^{-5}\Zs$}   &
		\colhead{$10^{-6}\Zs$}  
	}
	\startdata
	h1sn25    &$0.00$  &  $7.07\times10^{-6}$  & $5.01\times10^{-3} $  &   $5.06\times10^{-3} $    \\
	h1sn200  &$1.80\times 10^{-7}$  &  $1.81\times10^{-2}$  & $2.69\times10^{-2} $  &   $3.37\times10^{-2} $    \\
	h2sn25    &$0.00$  &  $0.00$  & $1.50\times10^{-4} $  &   $1.53\times10^{-4} $    \\
	h2sn200  &$3.32\times 10^{-8}$  &  $2.00\times10^{-3}$  & $4.29\times10^{-3} $  &   $6.02\times10^{-3} $  \\
	h3sn25    &$0.00$  &  $6.85\times10^{-6}$  & $2.29\times10^{-3} $  &   $2.32\times10^{-3} $    \\
	h3sn200  &$2.80\times 10^{-6}$  &  $2.53\times10^{-2}$  & $3.69\times10^{-2} $  &   $4.61\times10^{-2} $     
	\enddata
	\tablecomments{$\epsilon(z) = M_{\rm met}(z)/M_{\rm halo}$, $M_{\rm met}(z)$
		is the mass of metal-enriched gas at a minimum metallicity $z$, $M_{\rm halo}$ is the mass of  photoevaporated halo. }
	\label{tbl:met}
\end{deluxetable}

\subsection{Mixing in 3D}

\begin{figure*}
\plottwo{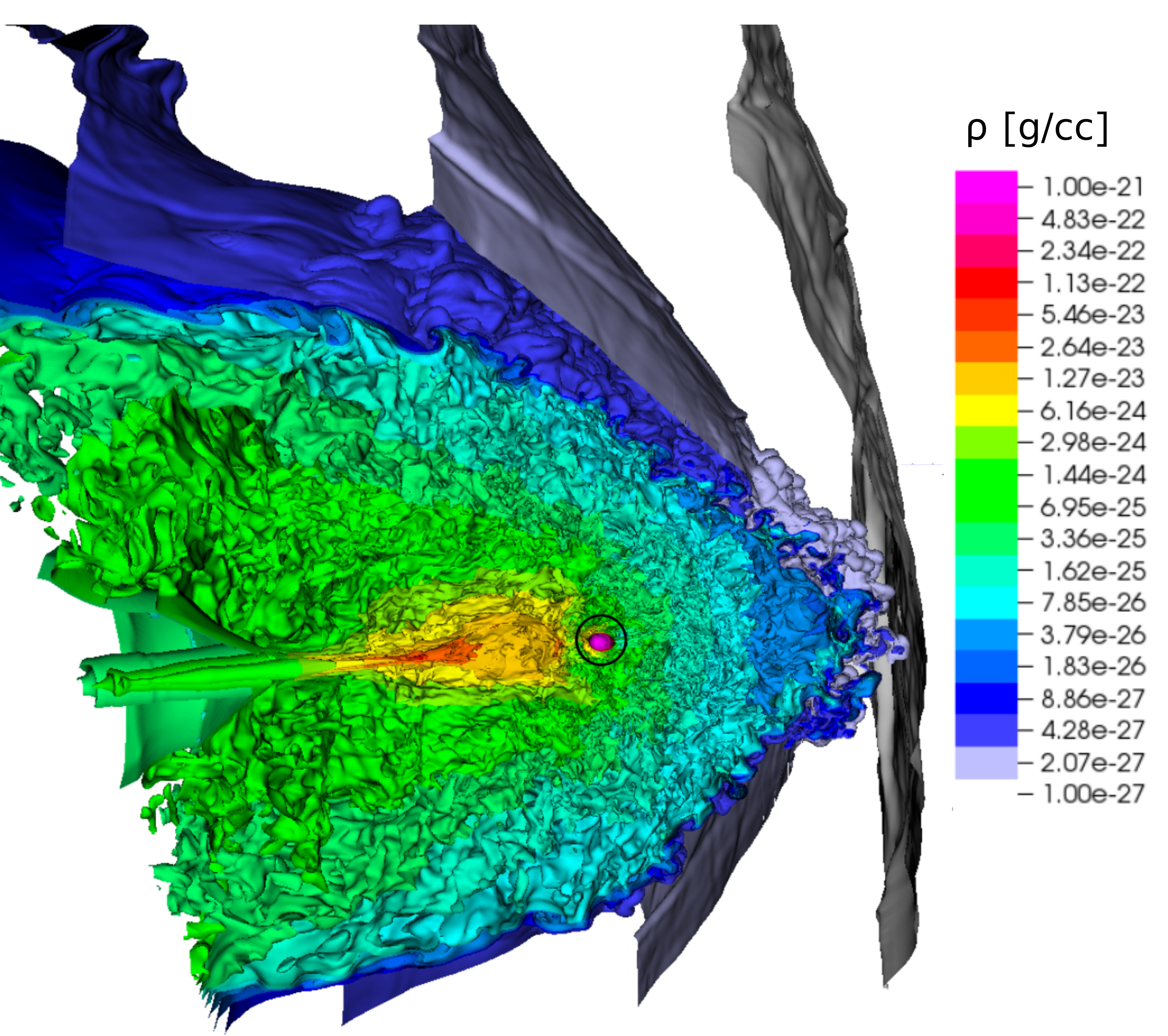}{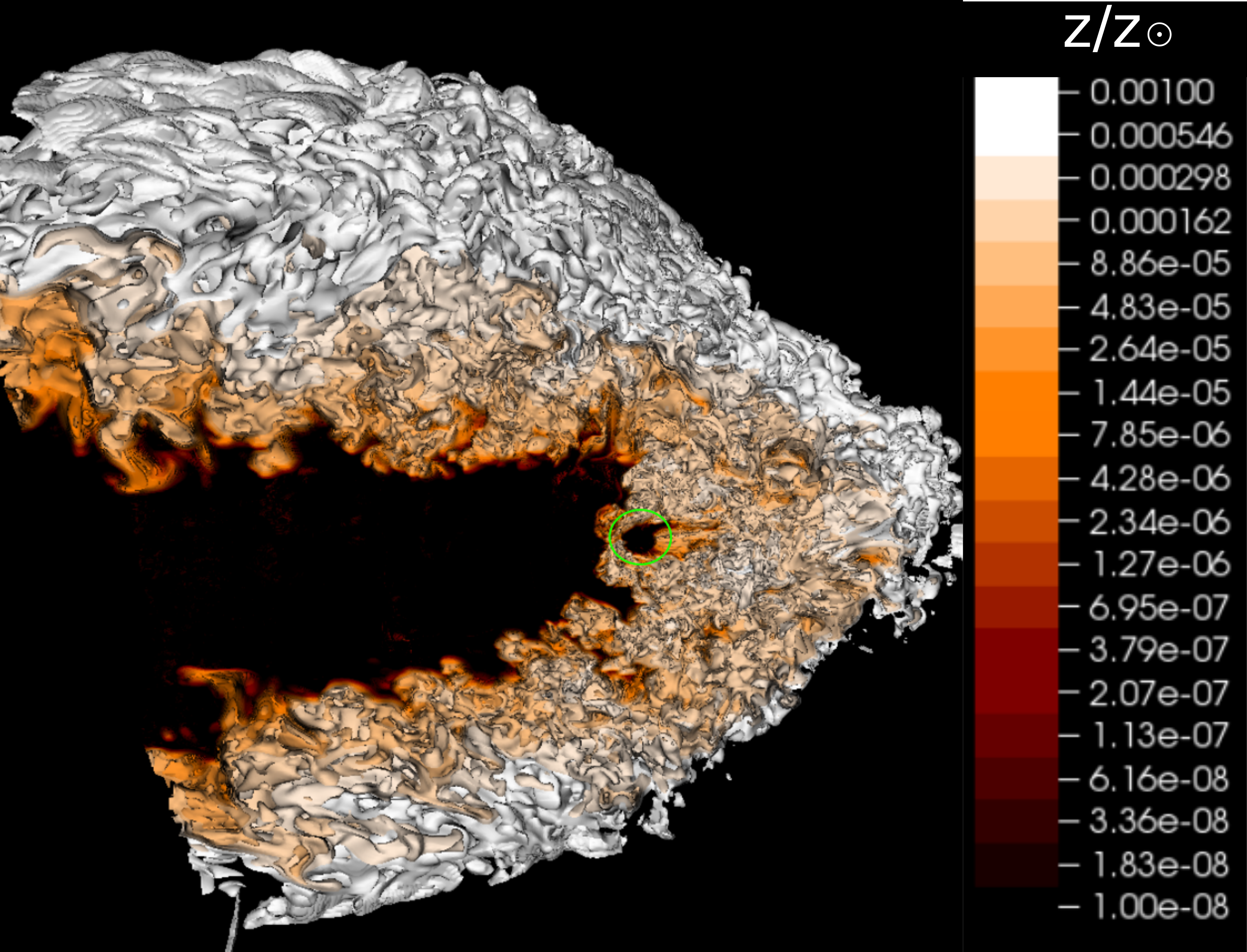}
\caption{Densities (left) and metallicities (right) in the halo in the h1sn200 run in 3D.  Colors 
show iso-surfaces of density after mixing. Turbulent flows are visible everywhere in zones of 
chemical enrichment.  The black circle marks the core of the halo, $r \sim 5$ pc. Metals also 
reach the core in this 3D run.} \vspace{0.1in}
\lFig{halo_den_3D}
\end{figure*} 

The development of turbulent cascades due to RT and KH instabilities is different in 3D and 2D, 
and it might be expected that mixing is more efficient in 3D because of the greater number of 
degrees of freedom of the flow. For example, the topology of I-front instabilities is quite different 
in 3D and they tend to devolve into turbulent flows much more efficiently \citep{wn08b,wn08a}.  
Here, we run the h1sn200 model in 3D to determine if mixing is enhanced or suppressed.  To 
set up this model we rotate the 2D profile of the halo and its ionized envelope around its axis of 
symmetry and align and center it along the $z$-axis of a 3D cartesian box in \CASTRO.  Had 
I-front instabilities developed in the H II region around the halo it would have been necessary to 
rerun the ZEUS-MP model in 3D as well because the structure of the instabilities would have 
been different, but none were present in this model. The simulation box is 512$^3$, with lower 
and upper boundaries at 0 and 250 pc along all three axes.  Metals flow in at the $z =$ 0 
boundary and outflow BCs are set on the other 5 boundaries.  Two levels of AMR refinement ($2^2$) 
in each dimension are used to achieve an effective resolution of 2048$^3$, or 0.122 pc.  The run required 1.5 
million CPU hours on 6000 cores on the Cray XC30 machine Aterui at CfCA at NAOJ. 

We show densities and metallicities for the halo at 10 Myr in Figure~\ref{fig:halo_den_3D}.  Both
images reveal significant turbulent mixing, and metals again reach the core of the halo.  Azimuthal 
averages of the metallicity in the 3D run are plotted with those from the 2D h1sn200 run in Figure~\ref{fig:met1}. 
The bulk kinetic energy of the flow is more efficiently dissipated by the additional 
degree of freedom in the turbulence in 3D, which enhances mixing at intermediate depths in the 
halo but results in fewer metals reaching the core, which is enriched to $\sim$ 10$^{-5}$ \Zs\ 
instead of a few 10$^{-4}$ \Zs. However, if dust forms in the metals it will still trigger the formation 
of Pop II stars there.   
 
\subsection{Comparison to Previous Models}

\djw{In contrast to \citet{cr08}, we find that metals from SNe can mix well with primordial halos when 
the evaporation of the halo by the progenitor star is taken into account because ionizing UV from
the star peel away the outer layers of the halo, partially exposing its core, and the outer layers of 
the halo are blown back toward the star and collide with debris from the explosion and mix 
efficiently with it.  Ejecta from the SN could only mix with the outer regions of the halo in \citet{
cr08} because they are not stripped away by the radiation, so only a few percent of its mass was 
chemically enriched (by KH shear instabilities in its outermost layers).  Preprocessing the halo
with UV radiation from the star is therefore crucial to how metals from the SN later mix with the 
halo. }

\citet{brit15} modeled the formation of a second-generation Pop II star in the debris of a Pop III 
SN in cosmological environments in the Enzo AMR code \citep{enzo}. In their model, a 40 \Ms\
Pop III star formed 200 pc from a 3 $\times$ 10$^{5}$ \Ms\ halo, partly evaporated it and then
exploded, with an energy of 10 B and a metal yield of 11.2 \Ms \citep{nom06}. The SN crashes 
into the halo at 6 Myr and enriches its core to $Z \sim$ 2 $\times$ 10$^{-5}$ \Zs, forming a Pop 
II star via dust cooling.  These results are consistent with our models because their initial UV  
fluxes and explosion energies were intermediate to those of our 25 and 200 \Ms\ stars.  Unlike 
our sn25 runs, metals from the SN in their simulation reached the core of the neighboring halo 
because it was exposed to higher UV fluxes over the life of the star (due to its larger mass and 
greater proximity to the halo) and because the explosion was more energetic and synthesized
more metals. But the metallicity to which the core was enriched was smaller than in our sn200
models because of the even greater UV flux, blast energy and metal yield of the 200 \Ms\ star.

\citet{rit16} examined the low-energy explosion of a 60 \Ms\ Pop III star inside a $\sim$ 10$^6$ 
\Ms\ halo that was not fully ionized by the star.  A dense neutral clump in the halo persisted and 
was later overrun by metals from the SN.  They found that the outer regions of the clump were 
only marginally enriched by metals, consistent with the mixing found in our 25 \Ms\ SN models.  
From the resolution of the simulation it was not clear if this clump would collapse to a Pop III 
star or a Pop II star, but this and the \citep{brit15} model exhibit the different degrees of mixing 
bracketed by our simulations.

\section{Discussion and Conclusion}

There is a region in the parameter space defined by the mass of the neighbor star, proximity 
to the halo, and the mass of the halo outside of which the SN cannot result in prompt Pop II 
star formation.  If the star is too close, too much of the halo will be evaporated, and whatever 
gas remains will be stripped from the core by the ram pressure of the SN ejecta.  The 
evaporative flow back towards the star will also be too diffuse to efficiently mix with the 
oncoming metals and form new stars.  If the star is too far away, too little of the halo is blown 
off and its core is mostly shielded from metals from the blast.  Not enough mass flows back 
towards the star to mix efficiently or fragment into new stars. The range of parameters in 
which second-generation Pop II stars may form as a direct result of the blast remains to be 
surveyed.  

Our models also indicate that even within this space the contamination of a halo with metals 
does not by itself guarantee the formation of Pop II stars, in contrast to what is usually 
assumed in semianalytical models or large-scale cosmological simulations of high-redshift 
star and galaxy formation.  The key in each case are the relative timescales of metal cooling 
and fragmentation in enriched gas and H$_2$ cooling in the pristine core because the 
formation of one type of star may preclude the formation of the other, depending on which 
happens first.  More simulations of halo-SN pairs with the cooling physics \djw{are needed} 
to resolve fragmentation and collapse are required to determine the branching ratios of Pop 
II and Pop III star formation (or both) in semianalytic models and simulations.

Although our halos are extracted from cosmological simulations, they are idealized in several
respects. Real halos have complex 3D geometries and are threaded by cosmological filaments
that are not present in our simulations. Collision of SN ejecta with the filaments could result in 
star formation in their outer layers, with some of these stars later being carried into the halo by 
the flow. Accretion from filaments and mergers with other structures also produce mild subsonic 
turbulence in the halo that could allow radiation from the star to reach to greater depths and 
promote mixing in the core (Smith et al. 2015; see also Greif et al. 2008).  Our models exclude 
turbulence due to mergers or accretion but capture turbulence due to RT and KH instabilities 
during the collision, and at a higher resolution than is currently achieved in cosmological 
simulations. That said, because we ignore turbulence in the halo prior to the collision our mixing 
results should be taken to be lower limits \citep[the same is true of][even without preprocessing 
of the halo with radiation]{cr08}.  How pre-collision turbulence enhances Pop II star formation in 
the halo will be investigated in future models, which may also include a prescription for turbulent
mixing on unresolved scales \citep{kl03,get09a}.

\djw{As discussed earlier, molecular hydrogen cooling in the relic H II region plays a role in 
mixing because it generally causes the gas to collapse to higher densities.  This effect is 
greatest in the earliest generations of stars because they build up a global LW background 
over time that dissociates H$_2$ and slows down cooling in these regions.  Reduced H$_2$ 
mass fractions would result in puffier, more diffuse relic H II regions that mix less efficiently 
with SN ejecta and suppress the formation of enriched stars.  Mixing and fragmentation in 
global LW backgrounds will be studied in future models.}

\djw{Another axis in the parameter space of mixing that is beyond the scope of this study is the initial degree of collapse of the halo
at the time of photoevaporation and collision with the ejecta.  Central densities in our halos
were $\sim 10^{5}-10^{6}$ cm$^{-3}$, so collapse was in progress but no stars had formed.  At even 
higher central densities the core of the halo is even more shielded from LW and ionizing UV
and metals cannot as easily reach it.  We therefore expect the contamination of the core to 
be reduced at later stages of collapse but not mixing at intermediate or outer layers.  If the
center of the halo rises to densities of $\gtrsim$ 10$^8$ cm$^{-3}$ before metals reach it the 
core fully molecularizes and collapses to a Pop III star with little or no enrichment.} 

\djw{Because our simulations do not include cooling due to metals we cannot say for certain 
when secondary enriched star formation will happen, only where it might occur due to mixing.  
Our models also lack cooling due to molecules and dust, which would also promote new star 
formation in mixed zones.}  Metals and molecules can cool gas down to the temperature of 
the CMB at densities of around $10^{4} \: {\rm cm^{-3}}$. Dust cooling occurs at much higher 
densities, regimes where the characteristic mass scale of fragmentation is much smaller due 
to the $\rho^{-1/2}$ dependence of the Jeans mass on density \citep{schn06}.  Numerical 
models indicate that Pop III SNe can produce up to several solar masses of molecules and 
dust \citep{cd09,cd10}, but the amount of dust that survives passage through the reverse 
shock in the SN ejecta remains unclear \citep{bian07,silv12}.  How early chemical enrichment 
schemes produce new stars will not be fully known until cooling due to all three species are 
implemented in simulations, which is now under development.

But cooling could also influence mixing prior to fragmentation by radiating away some of the bulk
kinetic energy of the flow.  It is clear from the difference in degree of mixing between PI SNe and 
less-energetic CC SNe that the loss this energy could dampen mixing and reduce the 
depths to which metals permeate the halo.  However, the mixing in our models is consistent with
that in cosmological simulations that do include these cooling processes, so it is unlikely that their
exclusion substantially alters our results.

A few dwarf galaxies have now been identified as candidates for 'one-shot' chemical enrichment, 
being contaminated by metals from just one Pop III SN at an early epoch and mostly isolated from 
mergers and accretion flows thereafter \citep{fb12}.  The elemental abundances and metallicity of 
Leo IV, for example, are a good fit to the yield of a single 10 \Ms\ Pop III CC SN \citep{sim10}. But 
our simulations suggest that Leo IV could not have been enriched by an external explosion of this 
mass and energy.  The UV flux from the progenitor would not have evaporated much of the halo 
prior to the explosion, and the energy of the blast would not have been sufficient to drive efficient 
mixing afterward (besides the fact that only metals in the solid angle subtended by the halo could 
have mixed with it). It is more likely that the elements in Leo IV came from internal enrichment, in 
which the weak UV flux of the Pop III star failed to ionize its host halo and the star exploded in a 
relatively compact and high-density H II region.  This scenario would lead to violent mixing, with 
few of the metals escaping the halo \citep{wet08a,ritt12}.

Finally, we have only considered one scenario in which metals from a SN enrich gas and trigger 
new star formation.  Many others can be imagined, such as collision of SN ejecta with the dense 
shell of a nearby relic H II region or with another SN remnant.  Even a single expanding SN front 
can crumple due to overstabilities and fragment, perhaps forming new stars. These and many 
other channels of chemically-enriched star formation are yet to be investigated.

\acknowledgements 

We would like to thank Mike Norman and Volker Bromm for useful discussions on the science 
in this paper and Ann Almgren, Mike Zingale, and Weiqun Zhang for their technical support 
with  \CASTRO. K.C. acknowledges the support of the EACOA
Fellowship of the East Asian Core Observatories Association and the hospitality of the Aspen 
Center for Physics, which is supported by National Science Foundation grant PHY-1066293.  
Work done at UCSC was supported by an IAU-Gruber Fellowship, the DOE HEP Program (DE-SC0010676) 
and the NASA Theory Program (NNX14AH34G).  D. J. W. was supported by STFC 
New Applicant Grant ST/P000509/1, and K. M. J. W. and D. J. W. were supported by the 
European Research Council under the European Community's Seventh Framework Programme 
(FP7/2007-2013) via the ERC Advanced Grant "STARLIGHT: Formation of the First Stars" 
(project number 339177).  All \CASTRO{} simulations were performed at NERSC and the CfCA 
at NAOJ.

\bibliographystyle{apjnew}
\bibliography{refs}

\end{document}